\documentclass[12pt,preprint]{aastex} 
 
\slugcomment{Not to appear in Nonlearned J. , 45. } 
 
\shorttitle{Pitch Angle of the Helical Magnetic Field} 
\shortauthors{Asada et al. } 
 
\begin{document} 
 
 
\title{MULTIFREQUENCY POLARIMETRY OF THE NRAO 140 JET: POSSIBLE DETECTION 
OF A HELICAL MAGNETIC FIELD AND CONSTRAINTS ON ITS PITCH ANGLE} 
 
 
\author{Keiichi Asada\altaffilmark{1,2}, Makoto Inoue\altaffilmark{1}, Masanori Nakamura\altaffilmark{3},  
Seiji Kameno\altaffilmark{1,4} and Hiroshi Nagai\altaffilmark{1,5}} 
\affil{National Astronomical Observatory of Japan, \\  
2-21-1 Osawa Mitaka Tokyo, 188-8588, Japan} 
\email{asada@vsop.isas.jaxa.jp} 
 
 
\altaffiltext{1}{National Astronomical Observatory of Japan} 
\altaffiltext{2}{Institute of Space and Astronautial Science, Japan Aerospace Exploration Agency, \\ 3-1-1 Yoshinodai, Sagamihara, Kanagawa, 229-8510, Japan} 
\altaffiltext{3}{Theoretical Astrophysics, Los Alamos National Laboratory} 
\altaffiltext{4}{Department of Physics, Faculty of Science, Kagoshima University} 
\altaffiltext{5}{Department of Astronomical Science, The Graduate University for Advanced Studies}

 
\begin{abstract}

We present results from multifrequency polarimetry of NRAO 140 using the Very Long Baseline Array. These 
observations allow us to reveal the distributions of both the polarization position angle and the Faraday rotation 
measure (RM). These distributions are powerful tools to discern the projected and line-of-sight components of 
the magnetic field, respectively. We find a systematic gradient in the RM distribution, with its sign being opposite 
at either side of the jet with respect to the jet axis. The sign of the RM changes only with the direction of the magnetic 
field component along the line of sight, so this can be explained by the existence of helical magnetic components 
associated with the jet itself. We derive two constraints for the pitch angle of the helical magnetic field 
from the distributions of the RM and the projected magnetic field; the RM distribution indicates that the helical 
fields are tightly wound, while that of the projected magnetic field suggests they are loosely wound around the jet 
axis. This inconsistency may be explained if the Faraday rotator is not cospatial with the emitting region. Our results 
may point toward a physical picture in which an ultra-relativistic jet (``spine'') with a loosely wound helical magnetic 
field is surrounded by a sub-relativistic wind layer (``sheath'') with a tightly wound helical magnetic field. 
 
\end{abstract} 
 
 



\section{Introduction} 
 
Jets from active galactic nuclei (AGNs) maintain well-collimated 
structure and can travel for more than 100 kpc. It has come to be 
known that the bulk Lorentz factors of these jets reach $\Gamma =$ 30 
in some cases (Kellermann et al. 2004). The mechanisms of formation, 
acceleration, and collimation of AGN jets are, however, 
still unclear. Magnetohydrodynamic (MHD) mechanisms are frequently 
invoked to model these jets. Magnetically driven outflows 
powered by a spinning black hole (Blandford \& Znajek 
1977) or an accretion disk (Blandford \& Payne 1982) have been 
widely discussed in terms of both the acceleration and collimation. 
MHD outflows originating in the AGN core, particularly 
those with a strong toroidal field encircling the collimated flow, 
can exhibit efficient acceleration and collimation. This continues 
over a scale of $10^3 - 10^4$ Schwarzschild radii, and the terminal 
Lorentz factors of MHD outflows have reached values of 
10-100 in recent numerical studies (Vlahakis \& K\"onigl 2004; 
Fendt \& Ouyed 2004; McKinney 2006). These theoretical pictures 
are qualitatively consistent with some VLBI observations 
(e.g., Junor et al. 1999; Horiuchi et al. 2006). In the theoretical 
models, the toroidal magnetic field plays an important role in 
the acceleration (through the magnetic pressure gradient force) 
and the collimation (through the magnetic tension force). Thus, 
the detection of helically twisted magnetic components associated 
with such jets is a crucial key to confirming the MHD models 
with observations. 
 
It is important to reveal the structure and strength of the magnetic 
fields in jets, although the available observational methods 
are limited. One powerful option is polarimetry. In the optically 
thin synchrotron emission from a nonrelativistic plasma, the polarization 
position angle (P.A.) is perpendicular to the projected 
direction of the magnetic field (the component perpendicular to 
the line of sight). In addition, the fractional polarization is closely 
related to the relative degree of ordering in the magnetic field. For 
this purpose, very long baseline polarimetry (VLBP) has been applied 
since the technique's early development. It has been reported 
that the P.A.'s tend to be parallel with respect to the jet 
axis for BL Lacertae objects whereas they tend to be perpendicular 
for quasars (Cawthorne et al. 1993). Jorstad et al. (2007), 
using multifrequency polarization observations, found a similar 
dichotomy but also suggested that it is not simply based on the 
optical classification of an object. The trend in BL Lac objects is 
sometimes explained by the compression due to strong shocks 
in the jets (Laing 1980; Hughes et al. 1989). Another possible 
explanation is that it is due to a toroidal magnetic field (Gabuzda 
et al. 2000) or the toroidal component of a helical magnetic field 
(Asada et al. 2002). However, the situation is not this simple, 
because the observed direction of the P.A. is not generally orthogonal 
to the projected direction of the magnetic field in the case of 
a relativistically moving, optically thin jet (Lyutikov et al. 2005). 
Neither is it simple to discriminate between these possibilities, 
since we have not taken advantage of the information derivable 
from the line-of-sight component of the magnetic field. 
 
Further possibilities for examining the structure of such magnetic fields have been provided by the capability of full polarimetry 
with the Very Long Baseline Array (VLBA). The line-of-sight component of the magnetic field can be probed by the distribution 
of the Faraday rotation measure (RM), since this is related to the electron density $n_{e}$ and the magnetic field component 
parallel to the line of sight ${\it B}_{\rm LOS}$ as RM $\sim \int_{\rm LOS} n_{e} {\it B}_{\rm LOS} dr$, where $\int_{\rm LOS} dr$  
represents integration along the line of sight. By combining the distributions of both the RM and the projected magnetic field,  
one can more reliably investigate the three-dimensional structure of the magnetic fields in parsec-scale jets. 
 
The first inference of the presence of a helical magnetic field 
structure from the RM distribution was for the 3C 273 jet, based 
on 5-8 GHz VLBA observations (Asada et al. 2002). Those 
observations revealed a gradient of the RM across the jet, which 
can be interpreted as indicating the presence of the toroidal component 
of a helical field (Asada et al. 2002). In the simple case 
where we are seeing such a field from the side (a viewing angle 
of 90$^{\circ}$), the sign of the line-of-sight component of the magnetic 
field will differ on the two sides of the jet as the field reverses 
direction. If the viewing angle decreases, the antisymmetric distribution 
will remain, with only an additional offset in the absolute 
value of the RM. Therefore, an RM gradient is expected 
in the presence of a helical magnetic field for arbitrary viewing 
angles, except 0$^{\circ}$. After this initial work, a similar RM gradient 
in 3C 273 was confirmed by independent observations (Zavala 
\&Taylor 2005; Attridge et al. 2005; Asada et al. 2008). The same 
kinds of gradients have been reported for several BL Lac objects 
as well (Gabuzda et al. 2004). 
 
In this paper we report observations of a bright quasar NRAO 140, which show an RM gradient across its jet,  
implying the presence of helical magnetic components. NRAO 140 is a distant quasar with a redshift, $z$, of 1. 263. 
If we assume ${\it H}_{0}$ = 73 km s$^{-1}$ Mpc$^{-1}$ and ${\it q}_{0}$ = 0. 5, 1 mas corresponds to 8. 4 pc. 
A value of 1 mas yr$^{-1}$ corresponds to 27. 4 {\it c}, and a superluminal motion of 11. 0 {\it c} was reported with VLBA observations at 15 GHz (Kellermann et al. 2004). 
 
The paper is organized as follows: In \S 2, we describe the details of the VLBA polarimetry observations and calibrations. 
In  \S 3,  we present polarization images of NRAO 140 jet and summarize several properties of the projected magnetic field and the RM. 
Discussion and conclusions are given in \S\S 4 and 5. 
 
\section{Observations and Data Reductions} 
 
The observations were carried out on 2003 January 13 using 
all 10 stations of the VLBA. We chose intermediate frequencies 
( IFs) of 4. 618, 4. 688, 4. 800, and 5. 093 GHz in the 5 GHz band 
and 8. 118, 8. 188, 8. 402, and 8. 593 GHz in the 8 GHz band. 
Each IF has an 8 MHz bandwidth. Both left- and right-circular 
polarizations were recorded at each station. The integration time 
was 55 minutes in each frequency band at various hour angles. 
We observed 3C 84 as a calibration source for the instrumental 
polarization and DA 193 for the polarization position angle. 
Calibrations were performed in the same manner as for our RM 
observations toward 3C 273, which are described in Asada et al. 
(2008). 
 
The instrumental polarizations of the antennas are determined for each IF at each band with 3C 84, using an AIPS task LPCAL. 
The polarization-angle offset at each station was calibrated using DA 193 obtained by VLA/VLBA Polarization Calibration Monitoring Program (Myers \& Taylor). The source was observed on 2002 Dec. 17 with VLA at 4. 8851 and 4. 8351 GHz and 8. 4351 and 8. 4851 GHz. 
Total flux of DA 193 measured with VLA is 5. 70 $\pm$ 0. 01 Jy at 5 GHz and 4. 98 $\pm$ 0. 01 Jy at 8 GHz, while those measured with VLBA are, in average over four IFs, 5. 42 $\pm$ 0. 24 Jy and 5. 01 $\pm$ 0. 21 Jy, respectively. 
In addition, total polarized flux measured with VLA is 55. 0 $\pm$ 2. 7 mJy at 5 GHz and 73. 1 $\pm$ 0. 1 mJy at 8 GHz, while those measured with VLBA are, in average over four IFs, 55 $\pm$ 15 mJy and 81 $\pm$ 3 mJy, respectively. 
Thus the time variation between our and VLA measurement would be small, and our calibration would be reasonable. 
 
Observing frequencies are slightly different between VLA and VLBA observations. 
We evaluated RM of DA 193 with VLA measurements, and we estimated PAs for each IFs by interpolation. 
In order to obtain the distributions of RM and projected magnetic field, we restored images at higher frequencies to match the resolution at the lowest frequency observation. The restored beam size was 2. 79 mas $\times$ 1. 28 mas with the major axis at a position angle of 0. 25$^{\circ}$. 
The distribution of RM was obtained by an AIPS task RM with polarization images at 4. 618, 5. 093, 8. 118, and 8. 593 GHz, with regions where the polarized intensity is three times greater than the r.m.s. noise in the polarized intensity. 
 
\section{Results} 
 
The P.A. distributions at 4. 618 and 8. 118 GHz are shown 
superposed on the distribution of total intensity in Figure 1. The 
core is presumably located at the northwest end of the jet, which 
extends to the southeast. The jet bends slightly toward the south 
at 10 mas from the core. We do not detect polarized flux from 
the core, at levels below 0. 3\% at 5 and 8 GHz. For the jet components, 
we detect significant polarized flux, at the 10\% level. 
The P.A.'s in the jet are nearly perpendicular to the jet axis but 
slightly reflect the curvature across the jet. The curvature is 
greater at 5 GHz than at 8 GHz. The difference in P.A.'s at the 
two frequencies can be explained by Faraday rotation, since 
the amount of rotation of the P.A. corresponds to $\lambda^{2}$. Therefore, 
nonuniform P.A. distributions result in a nonuniform RM 
distribution.

The distributions of the projected magnetic field and RM 
are shown superposed on the total intensity distribution at 
4. 618 GHz in Figure 2. The projected magnetic field runs 
roughly parallel to the jet axis, which is consistent with the usual 
trend seen in quasar jets (Cawthorne et al. 1993). Similarly to the 
case of the parsec-scale 3C 273 jet (Asada et al. 2002), we clearly 
see a gradient across the jet in the distribution of RM. In Figure 3, 
we show a cross section of the RM distribution along the line 
A-B from Figure 2, which is perpendicular to the jet axis. The 
 RMs differ by as much as 150 rad m$^{2}$, with an error less than 
40 rad m$^{2}$ and typically 20 rad m$^{2}$. The sign of the RM is 
positive at the southeast side of the jet and negative at the 
northwest side. The sign of the RM only depends on the direction 
of the line-of-sight component of the magnetic field. This 
change is naturally interpreted as being due to a reversal of this 
component. We note that similar reversals of RM sign on opposite 
jet sides have also been reported for QSOs 0745+241 and 
1652+398 (Gabuzda et al. 2004) and 3C 273 (Zavala \& Taylor 
2005).

The RM gradient is detected out to 8 mas from the core, or 
67 pc in projected distance. The viewing angle of the jet can be 
constrained from the Doppler effect as $\theta_{\rm max} = 2 \arctan{(1/\beta_{\rm app})}$, 
where $\theta_{\rm max}$ is the upper limit on the angle between the jet axis 
and the line of sight. Since an apparent motion $\beta_{\rm app}$ of 11c was 
detected from previous VLBA observations ( Kellermann et al. 
2004), this then gives an upper limit for the viewing angle 
of 10.4$^{\circ}$. Therefore, the linear extent of the jet is greater than 
350 pc, taking into account $\theta_{\rm max}$ $\sim$10.4$^{\circ}$. 
 
\section{Discussion} 
 
\subsection{Implications of the RM distribution} 
 
The RM gradient can be simply explained as the result of a 
helical magnetic field, as has been discussed for many AGN jets 
(Asada et al. 2002, 2008; Gabuzda et al. 2004; Zavala \& Taylor 
2005; Attridge et al. 2005). Similarly to 3C 273, we can estimate 
the direction of the toroidal component. The RM on the northeast 
side of the jet is positive, so the line-of-sight component of the 
magnetic field is directed toward us. On the other hand, that on 
the southwest side is negative, and the line-of-sight component of 
the magnetic field is directed away from us. Therefore, the field 
twists counterclockwise as seen from the core looking downstream 
through the jet. 
In the case of 3C 273, we discussed the possibility of detecting 
the longitudinal component of the helical magnetic field by 
using the offset of the RM (Asada et al. 2002). By combining 
the direction of the RM gradient and the estimated direction of 
the longitudinal component, we concluded that the orientation 
of 3C 273's helical field is that of a right-handed screw, reflecting 
the direction of rotation of the accretion disk or spinning 
black hole itself under the assumption that the helical magnetic 
field is wound up by the rotation. However, in NRAO 140 we 
could not detect any significant offset of the RM in its gradient. 
In the innermost regions of the jet, the RM is biased to the negative. 
If we assume that the offset can be ascribed to the longitudinal 
component of the helical magnetic field, we can uniquely 
predict the direction of the twisting of the helical magnetic field 
as being right-handed. If this field is wound by the rotation of 
the spinning black hole or the accretion disk, as suggested by 
the electromagnetic/MHD models (Blandford \& Znajek 1977; 
Blandford \& Payne 1982), it would appear that the northeastern 
half of the disk is approaching us and the southwestern half is 
receding. We identify the rotation direction of the accretion disk 
or black hole as clockwise as we see it, taking into account the 
configuration of the approaching jet. 
 
\subsection{Constraint for the pitch angle of the helical magnetic field} 
 
The RM gradient is a function of the pitch angle of the helical 
magnetic field and the viewing angle. Let us consider an observer 
simply seeing the field at a viewing angle of $\theta$. The line-of-sight 
component of the helical magnetic field, ${\rm {\it B}_{\rm LOS}}$, can be expressed 
as follows: 
\begin{eqnarray} 
{{\it B}_{\rm LOS} = {\it B}_{\psi} \sin \theta \sin \psi + {\it B}_{z} \cos \theta}. 
\end{eqnarray} 
where ${{\it B}_{\rm \psi}}$ and ${{\it B}_{z}}$ are the toroidal and longitudinal component of the helical magnetic field 
and ${\rm \psi}$ is the azimuthal angle, varying from -90$^{\circ}$ to 90$^{\circ}$. 
We show the line-of-sight component of the helical magnetic 
field across the jet at several pitch angles in Figure 4, assuming 
a viewing angle of 10.4$^{\circ}$, which is the same as the upper 
limit derived from the Doppler analysis. The sign of the line-of-sight 
component on one side is opposite to that on the other side 
when the pitch angle is smaller than  $\theta$, while the signs are identical 
when the pitch angle is larger than  $\theta$. The threshold for having 
both signs in the line-of-sight component is 
\begin{eqnarray} 
{{\it B}_{\rm \psi} \sin \theta > {\it B}_{z} \cos \theta}. 
\end{eqnarray} 
In the case of the NRAO 140 jet, we detect both positive and negative signs in the RM gradient. 
In addition, we derived an upper limit on the viewing angle, $\theta_{\rm max}$ = 10.4$^{\circ}$ from Doppler analysis. 
Thus, we can place a constraint on the pitch angle of the 
helical magnetic field in order to have both positive and negative 
signs in the RM gradient, since the sign of the RM can be 
changed by the direction of the line-of-sight component of the 
field. 
If we define the pitch angle of the helical magnetic field, $\chi$, as  ${\rm \tan \chi = {\it B}_{z}/{\it B}_{\psi}}$, 
\begin{eqnarray} 
{\rm \tan \chi < \tan \theta < \tan \theta_{\rm max}}. 
\end{eqnarray} 
This gives an upper limit for the pitch angle of the helical magnetic field of 10.4$^{\circ}$. 
 
On the other hand, a constraint on the pitch angle can also be 
derived from the direction of the projected magnetic field. Let 
us consider a simple case in which we are looking from the side 
at a helical magnetic field in an optically thin jet, following 
Asada et al. (2002). For a large pitch angle, the magnetic field 
runs almost parallel to the jet, and the P.A.'s on both near and far 
sides of the jet are nearly perpendicular to the jet. Then the P.A. 
integrated across the jet is also perpendicular to the jet, so that 
the resultant direction of the projected magnetic field is expected 
to be almost parallel to the jet. If instead the pitch angle is small, 
the resultant direction of the projected magnetic field is expected 
to be almost perpendicular to the jet. Therefore, a dichotomy in 
the direction of the projected magnetic field is expected, due to 
the vector accumulation of the incoherent-polarization radiation 
(see, e.g., Asada et al. 2002). Taking into account the viewing 
angle, the constraint for the projected magnetic field to be parallel 
to the jet can be expressed as 
 
\begin{eqnarray} 
{{\it B}_{z} \sin \theta >  {\it B}_{\rm \psi}}. 
\end{eqnarray} 
Again, if we define the pitch angle of the helical magnetic field by ${\tan \chi = {\it B}_{z}/{\it B}_{\rm \psi}}$, 
\begin{eqnarray} 
{\rm \tan \chi > \frac{1}{\sin \theta} > \frac{1}{\sin \theta_{\rm max}}}. 
\end{eqnarray} 
This gives a lower limit for the pitch angle of the helical magnetic field of 79.8$^{\circ}$. 
 
Thus, we have derived two constraints for the pitch angle 
of the helical magnetic field. 
One is  $\chi$ $<$ 10.4$^{\circ}$, and the other is $\chi$ $>$ 79.8$^{\circ}$. 
These are obviously inconsistent, an inconsistency 
that may come from a difference between the region emitting 
high-energy particles and that corresponding to the Faraday rotation, 
that is, a Faraday screen. In fact, a reasonably strong fractional 
polarization of 10\% is detected with an RM of 100 rad m$^{2}$. 
For a simple slab model of internal Faraday rotation, the fractional 
polarization is expected to be smaller than 1.7\% at 5 GHz 
with an RM of 100 rad m$^{2}$. Therefore, this would exclude internal 
Faraday rotation, and we suggest that a layer or sheath 
surrounding the emitting jet is responsible for the Faraday rotation 
(Inoue et al. 2003).

Such a spine-sheath structure has been strongly supported in 
several cases in terms of the amount of Faraday rotation being 
greater than 90$^{\circ}$ ( Inoue et al. 2003; Gabuzda et al. 2004; Zavala 
\& Taylor 2005). A rapid time variation of the RM gradient in 
the 3C 273 jet also implies that MHD-driven AGN jets may 
have multiple, layered components (Asada et al. 2008). From a 
theoretical point of view, the spine may correspond to the ultra-relativistic 
outflows, with high Lorentz factors of $\Gamma \geq 10$. This 
idea originates in the fact that the electromagnetic (EM) energy 
strongly dominates the matter energy in the outflows produced 
by a spinning black hole (Blandford \& Znajek 1977). Thanks to 
the lack of matter inertia, the magnetic field has a large pitch 
angle (the local Alfv\'en speed may almost reach the speed of 
light, and an inertial back-reaction of the plasma on the poloidal 
field may not occur efficiently in the EM-dominated regions). This 
physical picture has been observed in recent general relativistic 
degenerate electrodynamic and MHD simulations (Komissarov 
2001, 2004). However, the sheath may correspond to the sub-relativistic 
or mildly relativistic outflows, with low Lorentz factors 
of a few, and arise from the accretion disk (Tsinganos \& 
Bogovalov 2002). The matter can be accelerated by the Lorentz 
force, and the flows reach a superfast magnetosonic regime far 
from the disk. The flow may finally become matter dominated 
(kinetic plus internal energy), with a low pitch angle (Blandford 
\& Payne 1982; Vlahakis \& Ko\"nigl 2004). Thus, a picture 
emerges in which an ultra-relativistic ``jet'' (the spine) with a 
loosely wound helical field, perhaps surrounded by mildly relativistic 
``winds'' (the sheath) with a tightly wound helical field, 
may be a common structure in AGN jets, which is essentially consistent 
with our results. 
 
On the other hand, in the case of M87 a Hubble Space Telescope 
({\it HST}) observation provides us with angular resolution of 
0.2 arcsec. This corresponds to a linear resolution of 15 pc, which is 
similar to that of our VLBA observations of NRAO 140. Based 
on this {\it HST} observation, Perlman et al. (1999) found that the 
M87 jet appears narrower in the optical than in the radio. This 
suggests that for arcsecond-scale jets, the jet consists of a 
highly relativistic spine and a lower velocity sheath; the highly 
relativistic spine may be responsible for the optical emission, 
and the lower velocity sheath may be responsible for the radio 
emission. If the slow sheath observed as an RM gradient exists 
in arcsecond-scale jets, this would indicate that there is another 
sheath wrapped around the highly relativistic spine and lower 
velocity sheath, suggesting that a stratified flow structure is more 
likely than a simple two-component model. A similar tendency 
has been reported for 3C 273 as well (Jester et al. 2005). 
 
\section{Conclusions} 
 
In order to discuss the three-dimensional configuration of the 
magnetic fields in parsec-scale quasar jets, we have performed 
multifrequency VLBA polarimetry toward NRAO 140. We revealed 
the distributions of both the projected component of the 
magnetic field and the rotation measure. We find a systematic 
gradient in the RM distribution similar to that seen in the 3C 273 
jet. Furthermore, the sign of the RM is opposite on either side of 
the gradient. This is presumably due to a change of direction of 
the magnetic field and is naturally explained by a helical field 
structure. Using the properties of the RM gradient and the viewing 
angle, we derive a constraint for the pitch angle of the helical 
magnetic field in the layer or the sheath of the jet of $\chi$ $<$ 10.4$^{\circ}$. 
On the other hand, from the direction of the projected magnetic 
field and the viewing angle, we derive another constraint for the 
pitch angle in the spine of the jet of $\chi$ $>$ 79.8$^{\circ}$. Therefore, we 
expect that in NRAO 140 an ultra-relativistic jet (or ``spine'') with 
a loosely wound helical magnetic field is surrounded by a layer 
(or ``sheath'') with a tightly wound helical magnetic field.

{\bf Acknowledgments:} 
The authors thank E. S. Perlman for a critical reading of this paper and valuable comments as a referee. 
The VLBA and VLA are operated by the National Radio Astronomy Observatory (NRAO),  
a facility of the National Science Foundation, operated under cooperative agreement by Associated Universities, Inc.

 

 

\clearpage 
 
 

 
\begin{figure} 
\plottwo{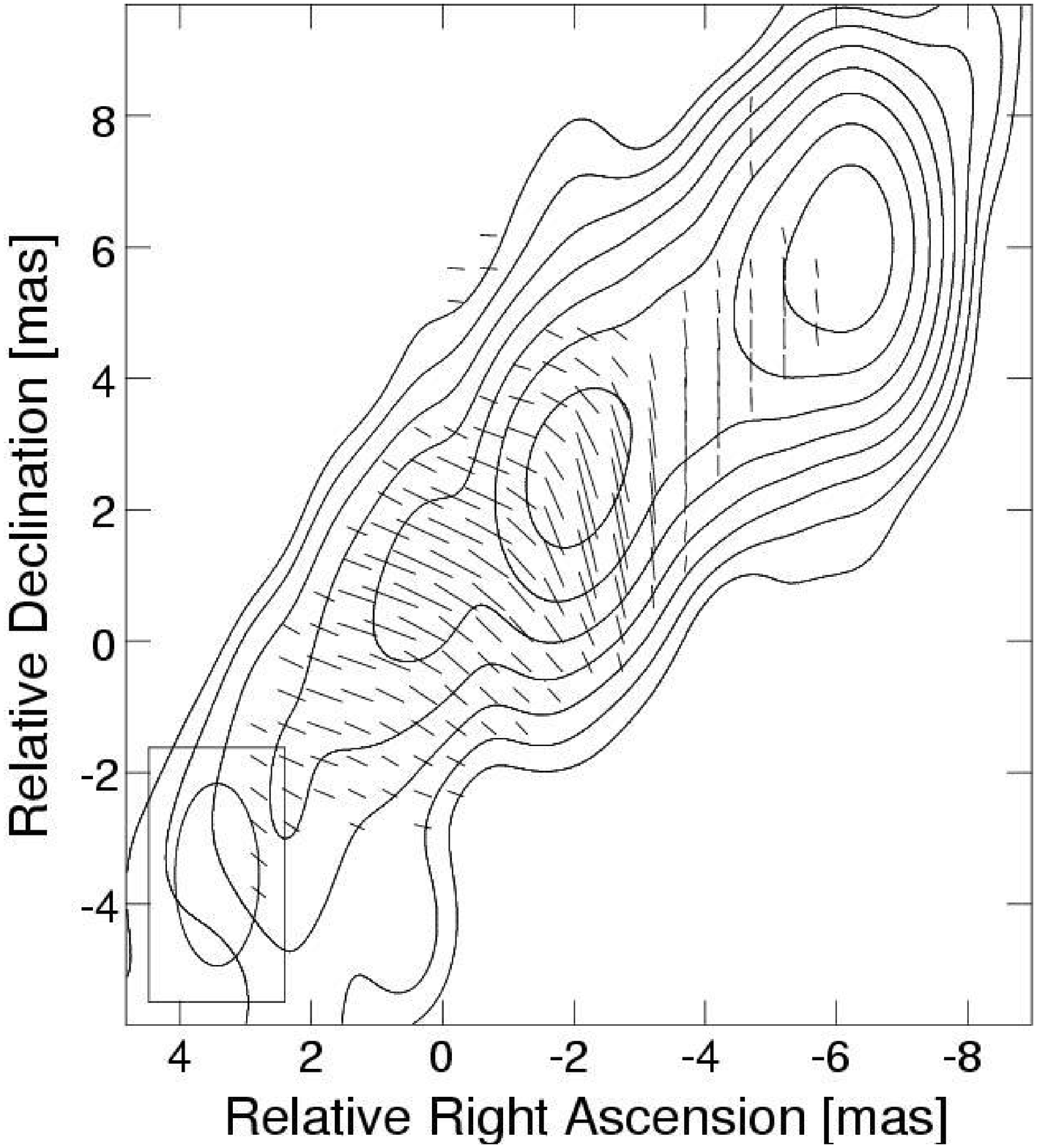}{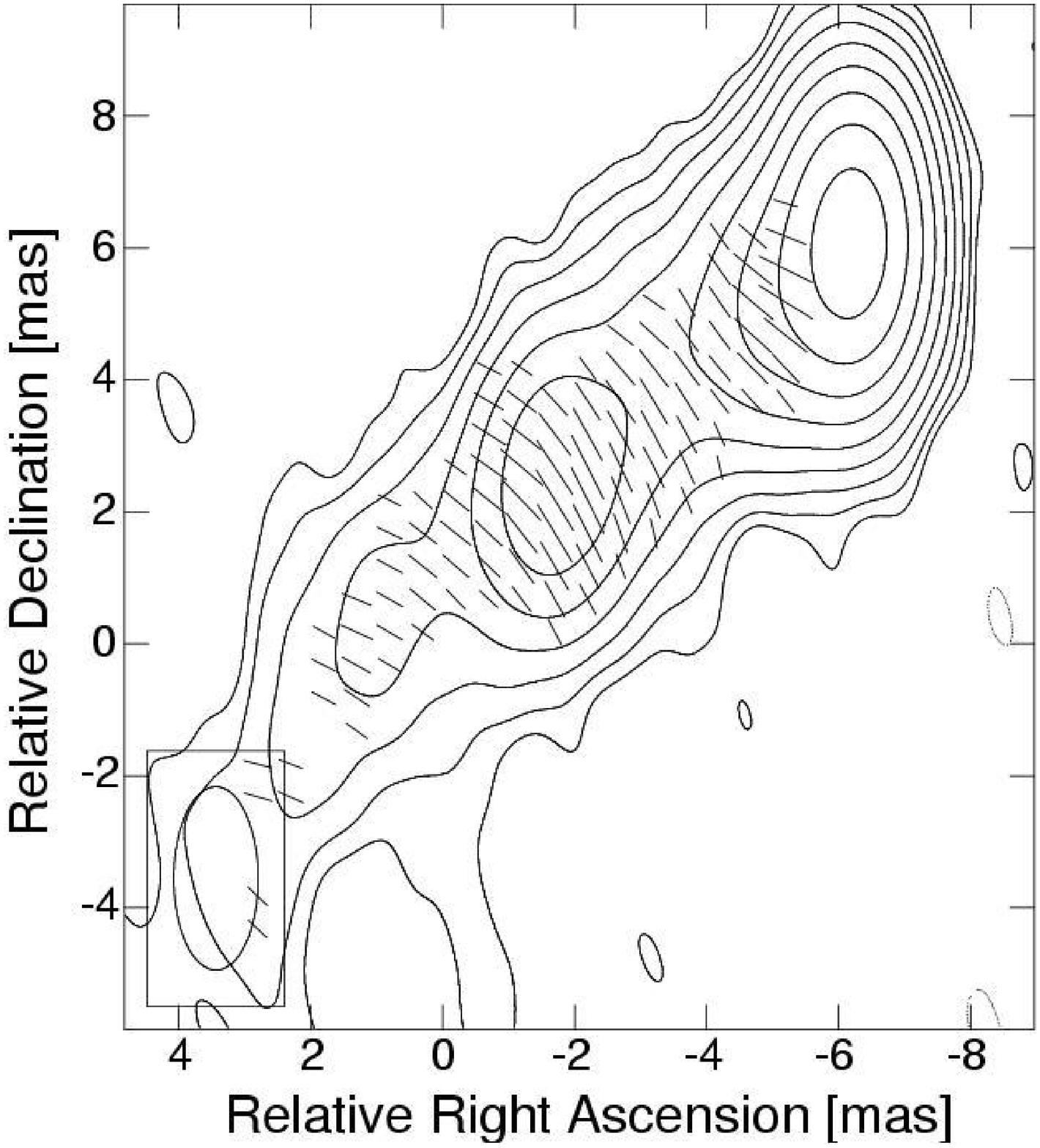} 
\caption{(left) Distribution of polarization position angle (bar) superposed on the contour images of the total intensity at 4. 618 GHz. (right)  
 Distribution of polarization position angle (bar) superposed on the contour images at 8. 118 GHz. Contours are plotted at -1, 1, 2, 4, 8, 16, 32, 64, 128, 256, 512 and 1024 $\times$ three times the r. m. s. noise of  
that of the total intensity at each frequency. The synthesized beam size is restored by the 4. 618 GHz beam of 2. 79 mas $\times$ 1. 28 mas with the major axis at a position angle of 0. 25$^{\circ}$. Image at 8. 118 GHz is restored to the same resolution of 4. 618 GHz. 
\label{POLC}} 
\end{figure} 
 
\begin{figure} 
\plottwo{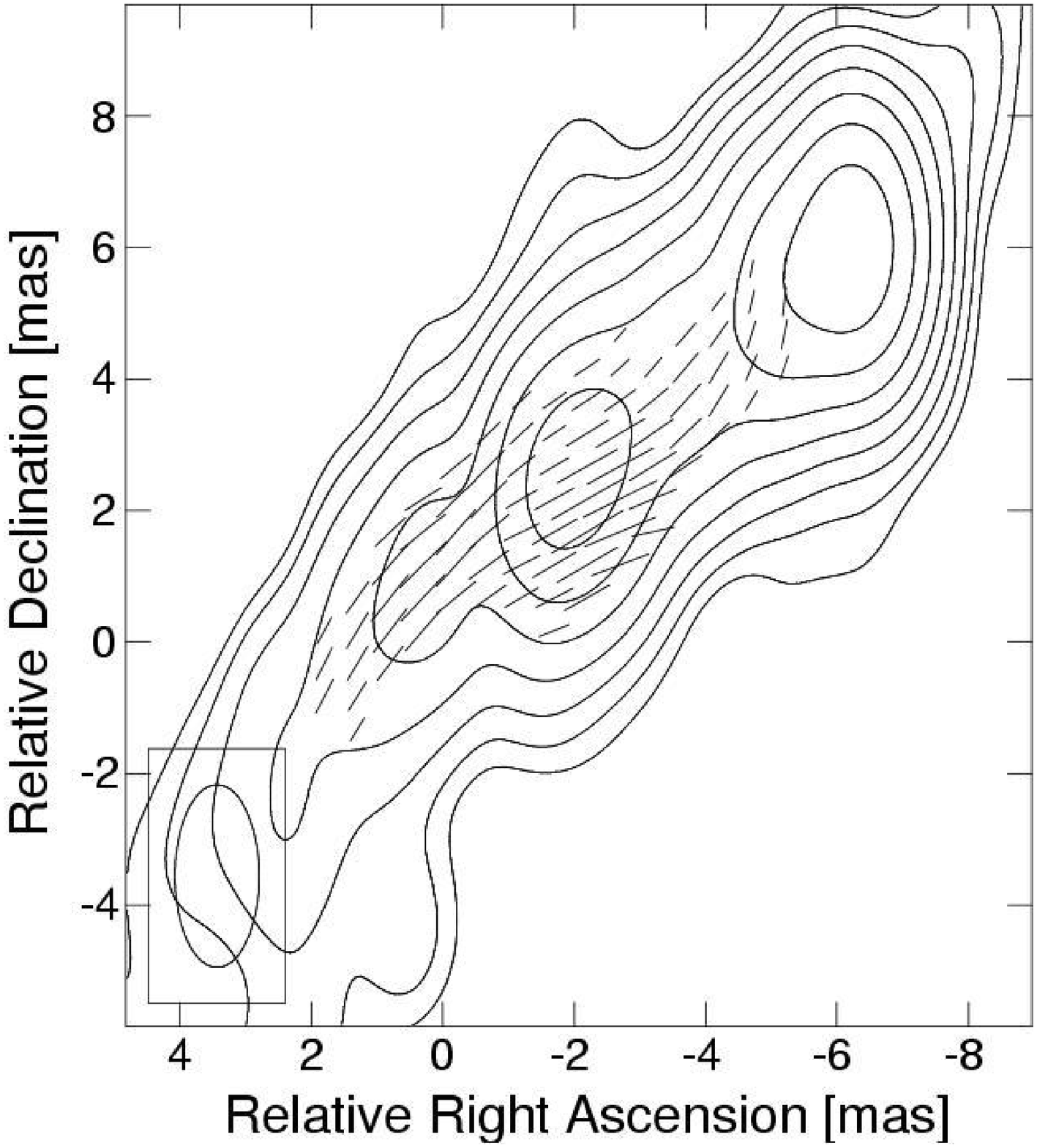}{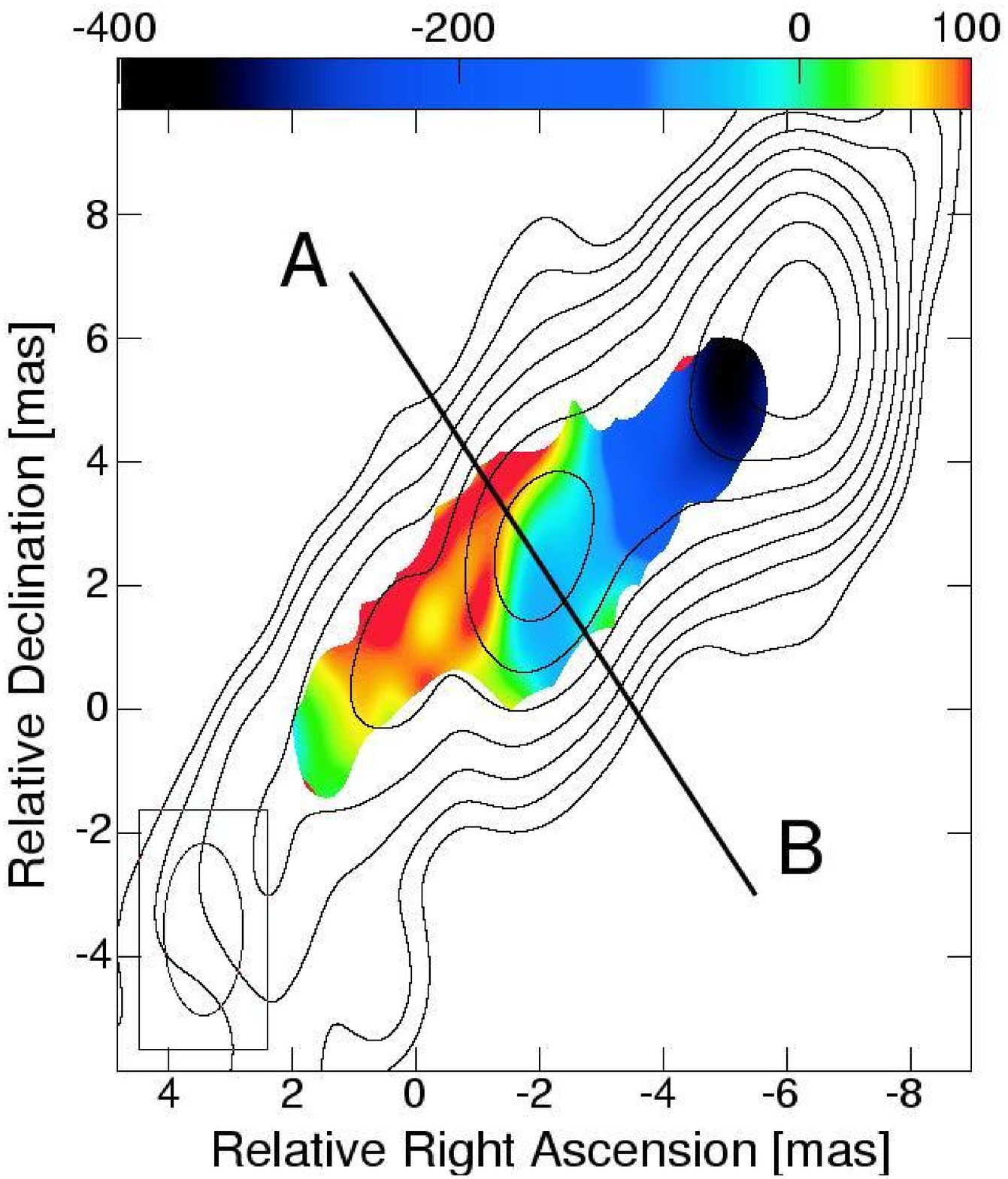} 
\caption{(left) Distribution of the projected component of the magnetic field (bar) superposed on the contour images of the total  
intensity at 4. 618 GHz. Contours are plotted at -1, 1, 2, 4, 8, 16, 32, 64, 128, 256, 512 and 1024 $\times$ three times the r. m. s. noise  
of that of the total intensity at 4. 618 GHz. The synthesized beam size is restored by the 4. 618 GHz beam of 2. 79 mas $\times$ 1. 28 mas  
with the major axis at a position angle of 0. 25$^{\circ}$. (right) Distribution of RM (color scale) superposed on the contour  
images of the total intensity at 4. 618 GHz. The RMs are plotted in the region where the polarized intensity is greater than three times the r. m. s. noise in the polarized intensity. 
\label{ RM _B}} 
\end{figure} 
 
\begin{figure} 
\epsscale{.80} 
\plotone{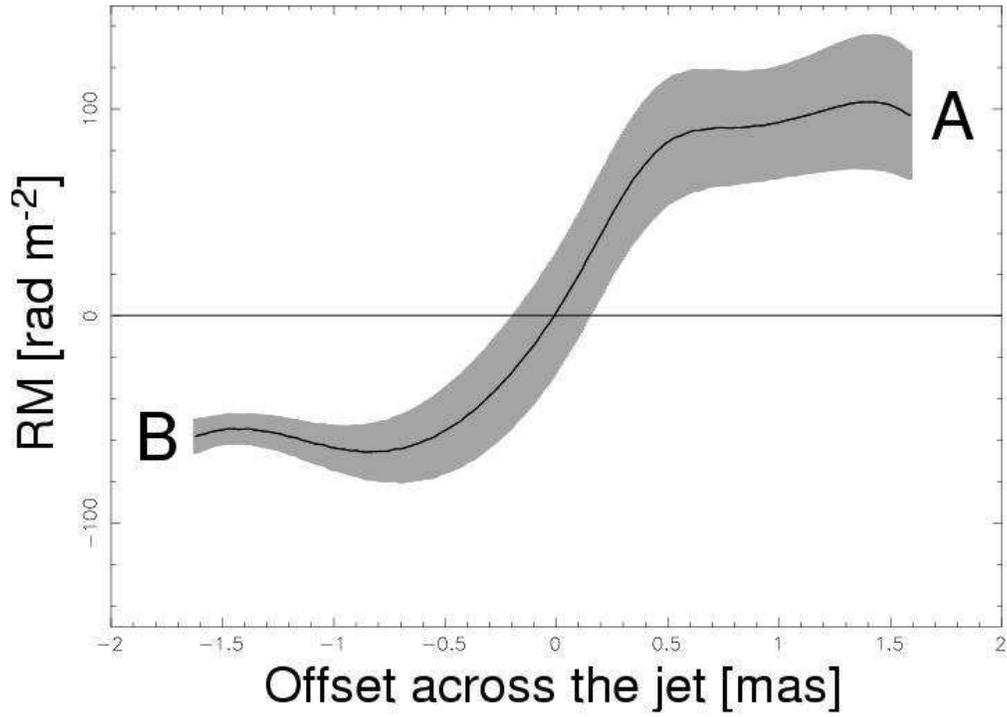} 
\caption{Cross sections of the RM distribution along the line A--B, derived using the AIPS task SLICE. The shaded area along the curved line of the RM indicates the standard deviation (1 $\sigma$) in the RM. The profile of the RM distribution is anti-symmetric with respect to the central axis of the jet. 
\label{CS}} 
\end{figure} 
 
\begin{figure} 
\epsscale{.80} 
\plotone{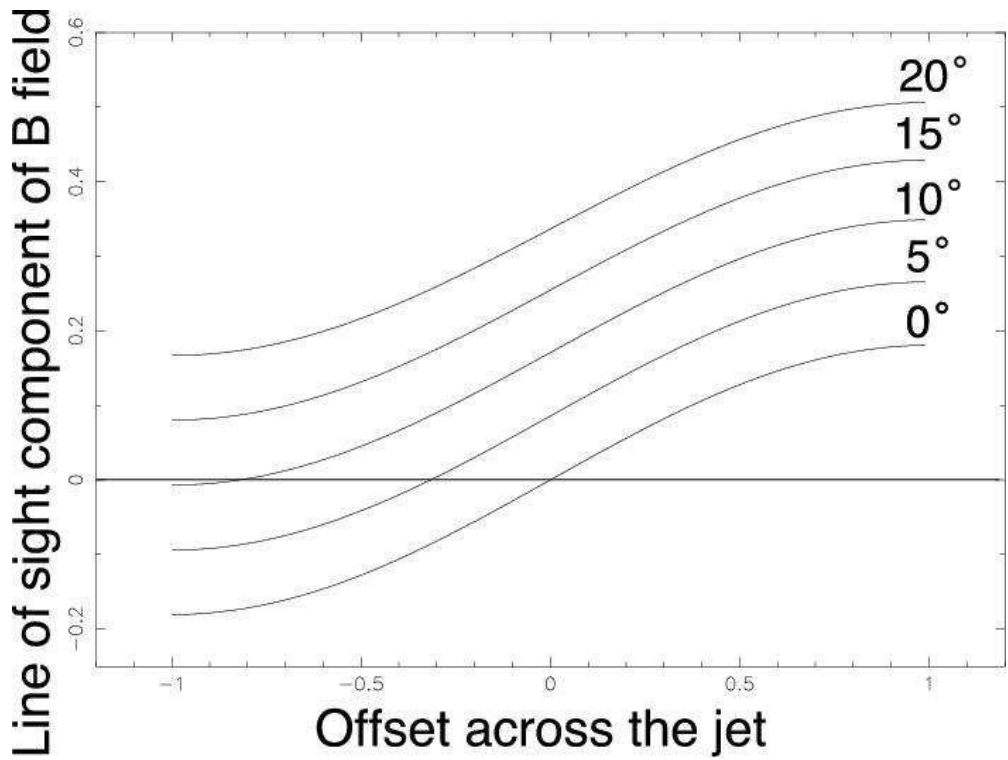} 
\caption{Line-of-sight component of the helical magnetic field with pitch angle at 0$^{\circ}$, 5$^{\circ}$, 10$^{\circ}$, 15$^{\circ}$ and 20. $^{\circ}$ The viewing angle of the helical magnetic field is 10.4$^{\circ}$, which is same as the upper limit of the viewing angle derived from the Doppler analysis. Vertical and horizontal scales are normalized by the strength of the magnetic field and the width of the jet, respectively. 
\label{LOS}} 
\end{figure}

\end{document}